\documentclass[%
 reprint,
superscriptaddress,
 amsmath,amssymb,
 aps,
]{revtex4-1}

	\usepackage{bm}
	\usepackage{braket}
	\usepackage{dsfont}

 	\usepackage{graphicx}
	\usepackage[caption=false]{subfig}
	\usepackage[export]{adjustbox}

	\newcommand{\vect}[1]{\boldsymbol{#1}}		
	\newcommand{\op}[1]{\hat{\boldsymbol{#1}}}	
	

\begin{document}

\title{Negative differential resistance in Van der Waals heterostructures due to moir\'{e}-induced spectral reconstruction}

\author{Damien J. Leech} 
\email{D.J.Leech@bath.ac.uk}
\affiliation{Department of Physics, University of Bath, Claverton Down, Bath BA2 7AY, United Kingdom}
\author{Joshua J. P. Thompson}
\email{J.J.P.Thompson@bath.ac.uk}
\affiliation{Department of Physics, University of Bath, Claverton Down, Bath BA2 7AY, United Kingdom}
\author{Marcin Mucha-Kruczy\'{n}ski}
\affiliation{Department of Physics, University of Bath, Claverton Down, Bath BA2 7AY, United Kingdom}
\affiliation{Centre for Nanoscience and Nanotechnology, University of Bath, Claverton Down, Bath BA2 7AY, United Kingdom}

\date{\today}

\begin{abstract}

Formation of moir\'{e} superlattices is common in Van der Waals heterostructures as a result of the mismatch between lattice constants and misalignment of crystallographic directions of the constituent two-dimensional crystals. Here we discuss theoretically electron transport in a Van der Waals tunnelling transistor in which one or both of the electrodes are made of two crystals forming a moir\'{e} superlattice at their interface. As a proof of concept, we  investigate structures containing either aligned graphene/hexagonal boron nitride heterostructure or twisted bilayer graphene and show that negative differential resistance is possible in such transistors and that this arises as a consequence of the superlattice-induced changes in the electronic density of states and without the need of momentum-conserving tunnelling present in high-quality exfoliated devices.  We extend this concept to a device with electrodes consisting of aligned graphene on  $\alpha - \text{In}_2 \text{Te}_2$ and demonstrate  negative differential resistance peak-to-valley ratios, $\sim 10$.
\end{abstract}

\maketitle

\section{Introduction}
	
\begin{figure*}
\subfloat{  \includegraphics[clip,width=2\columnwidth,valign=t]{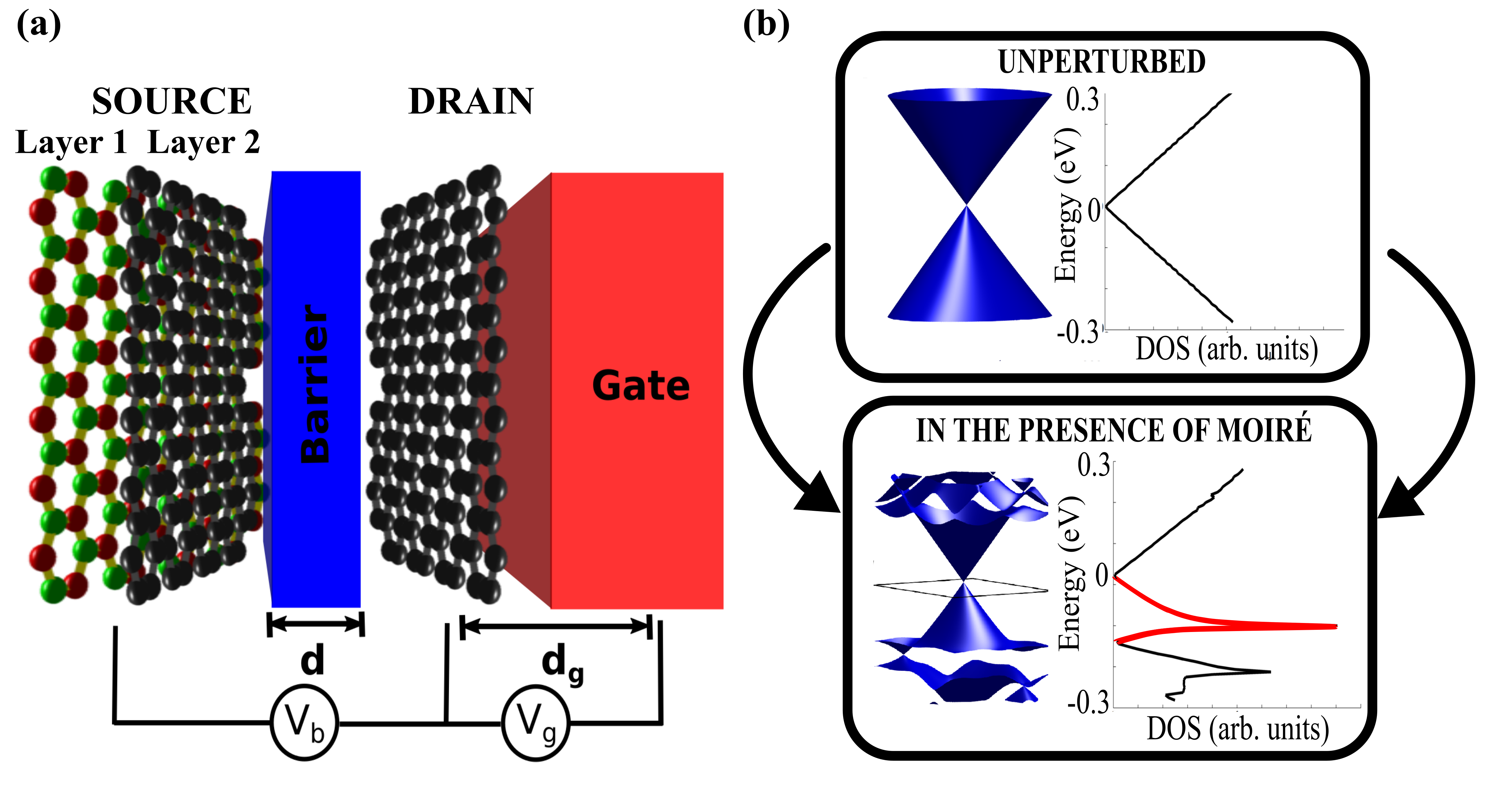} \label{fig:device}}
\caption{(a) Schematic of the tunnelling device with a source made of an aligned graphene/hBN heterostructure. Also shown are the distances $d$ and $d_{g}$, as well as contacts for voltages $V_{b}$ and $V_{g}$. (b) The conical bands and density of states of unperturbed monolayer graphene (top) and corresponding graphene minibands and density of states of an aligned graphene/hBN heterostructure (bottom). In the miniband spectrum, the black lines indicate the boundaries of a rhombic superlattice Brillouin zone. The Van Hove singularity in the density of states of the heterostructure is highlighted in red.}
\label{fig:DeviceAndDOSMiniBand}
\end{figure*}

The phenomenon of negative differential resistance (NDR) is a striking example of nonlinearities in physics --- within a certain region of the current/voltage characteristic of a device, increase of applied voltage leads to decrease of the output current. In the first solid state device displaying NDR, the Esaki diode \cite{esaki_physrev_1958}, this effect arises because increasing bias voltage modifies alignment of the occupied and empty electronic states in the source and drain electrodes separated by a tunnelling barrier.  At zero and large bias, either due to the lack of occupied states at the source or empty states at the drain, this alignment prohibits flow of current. In contrast, within a certain bias window in between these two cases, the positioning of energy levels allows electrons to tunnel through the barrier.

More recently, negative differential resistance was observed in Van der Waals (VdW) heterostructures of two-dimensional atomic crystals as a result of momentum-conserving electron tunnelling through an atomically thin barrier \cite{britnell_natcom_2013, mishchenko_natnano_2014, fallahazad_nanolett_2015, kang_ieee_2015, kkim_nanolett_2016, wallbank_sci_2016}. Due to the high quality of the crystals produced by mechanical exfoliation and the atomically sharp interfaces in the assembled VdW-coupled stack \cite{haigh_natmat_2012}, the requirement to match both energy and momentum of the initial and final states leads to a peak in the tunnelling current as applied voltages tune the source and drain to a particular band alignment. However, exfoliation, while providing state-of-the-art materials and devices, is not a scalable fabrication method. At the same time, materials produced by other methods such as chemical vapour deposition do not achieve the quality necessary to observe momentum-conserving tunnelling and seemingly NDR \cite{roy_applphyslett_2014, lee_apl_2014}.

Here, we show theoretically that NDR can be achieved in VdW heterostructures without momentum-conserving tunnelling. Instead, we exploit modifications of the electronic band structure of such heterostructures due to the interplay between lattice constants as well as misalignment of the crystallographic axes of two neighbouring layers, which lead to the formation of a superlattice at the interface. This superlattice is commonly referred to as the moir\'{e} pattern and is unique to VdW heterostructures in which, due to the Van der Waals coupling between different materials, lattice matching is not necessary for the whole structure to be stable. Crucially, formation of the moir\'{e} superlattice is often accompanied by reconstruction of the electronic band structure as the moir\'{e} periodicity folds the dispersion into minibands. This results in opening of mini gaps at the boundary of the superlattice Brillouin zone and appearance of Van Hove singularities in the electronic density of states \cite{yankowitz_natphys_2012, wang_natphys_2016, shi_natphys_2014, yang_nanolett_2016, li_natphys_2010, yan_physrevlett_2012, tan_acsnano_2016, ykim_nanolett_2016, pierucci_nanolett_2016, koren_natnanotech_2016, brihuega_prl_2012, lee_science_2016}. We simulate the tunnelling current in two common VdW heterostructures in which the source electrode is either (1) monolayer graphene highly aligned with underlying hexagonal boron nitride (hBN) or (2) twisted bilayer graphene and show that moir\'{e}-induced spectral reconstruction can result in negative differential resistance. We then study a more complex device with both the source and drain electrodes made of a moir\'{e}-forming stack. We show that in the case of graphene/$\alpha - \text{In}_2 \text{Te}_2$ electrodes, NDR an order of magnitude larger than for the two previous architectures is possible, suggesting that design of new Van der Waals interfaces can provide a way to engineer current characteristics of tunneling junctions, including NDR.

\section{Device Architecture}

A general schematic of our VdW-based tunnelling transistor is shown in Fig. \ref{fig:device}. It comprises two electrodes, referred to as source and drain, separated by a thin tunnelling barrier. A bias voltage $V_{b}$ is applied between the two electrodes while a gate controls an additional voltage $V_{g}$. Here,  the source electrode consists of two layers arranged in such a way that layer 1, further from the barrier, generates a long-wavelength periodic potential for electrons in layer 2, perturbing their electronic states. In structures involving two-dimensional atomic crystals, such periodic potentials arise naturally as a consequence of different lattice constants of the neighbouring materials as well as any misalignment, $\theta$, between their respective crystallographic axes. This leads to the formation of superlattices visually represented by moir\'{e} patterns seen for examples in scanning tunnelling microscopy measurements \cite{yankowitz_natphys_2012, wang_natphys_2016, yang_nanolett_2016, li_natphys_2010, yan_physrevlett_2012}. While the impact of the moir\'{e} perturbation depends on the atomic composition of the two layers and details of the geometry, in many systems the additional potential leads to Bragg scattering of the electrons and folding of the electronic spectrum into the superlattice Brillouin zone (sBZ) accompanied by opening of minigaps along its boundary \cite{yankowitz_natphys_2012, wang_natphys_2016, shi_natphys_2014, yang_nanolett_2016, li_natphys_2010, yan_physrevlett_2012, tan_acsnano_2016, ykim_nanolett_2016, pierucci_nanolett_2016, koren_natnanotech_2016, brihuega_prl_2012, lee_science_2016}, as indicated schematically in Fig. 1b for the case of graphene on hexagonal boron nitride. As a result of such a spectral reconstruction, the electronic density of states (DOS) is strongly modified - a fact crucial to the functioning of our device as, for a thin tunnelling barrier, the current $I$ is sensitive to the source and drain DOS \cite{moffat}, $\rho_{s}$ and $\rho_{d}$, respectively,

\begin{align}
\begin{split}\label{CurrentEq}
    I \!=\! \frac{2g \pi e}{\hbar} & \int T (\epsilon) \rho_{s} (\epsilon) \rho_{d} (\epsilon \! - \! \Delta ) \\ &  \times [f(\epsilon \! - \! \mu_{s}) \! - \! f(\epsilon \! - \! \Delta \! - \! \mu_{d})] d\epsilon,
\end{split}
\end{align}
where the energy $\epsilon$ is measured from the source charge neutrality point, $\mu_s$ and $\mu_d$ determine the energy distance between the chemical potential and the charge neutrality points in the source and drain  electrode, respectively, $\Delta$ is the shift between the source and drain neutrality points  so that  $\mu_s$ and $\mu_d+ \Delta$ are the chemical potentials in the corresponding electrodes, $T$ is the transmission coefficient, $f(\epsilon)$ is the Fermi-Dirac distribution (here, we take the low-temperature limit) and $g$ takes into account additional degeneracies of the electronic states (here, spin and valley).  
Without the additional moir\'{e} modulation, the DOS of the electrodes are usually varying slowly (linearly for monolayer graphene, constant for quasi-free electrons in 2D) and no NDR is observed \cite{britnell_sci_2012, roy_applphyslett_2014, lee_apl_2014} in the absence of momentum-conserving tunnelling - this might be the case for devices of insufficient quality or large misalignments between the crystallographic directions of the electrodes. Note that, if the momentum-conserving tunnelling processes  become important, their contribution cannot be larger than the currents discussed here, because ultimately the number of tunnelling electrons is set by the corresponding DOS whereas momentum conservation adds an additional constraint that is only fulfilled by some of them. 

In order to relate the energies $\Delta$, $\mu_{s}$ and $\mu_{d}$ to the applied voltages $V_{b}$ and $V_{g}$, we use the electrostatic relations

\begin{align} \label{Electrostatics}
    & V_{b} = \frac{1}{e}\left( \mu_{s} - \mu_{d} - \Delta \right), \\
    & \Delta = \frac{e^{2}d}{\varepsilon_{0} \tilde{\varepsilon}}\left( n(\mu_{d}) + V_{g} \frac{\varepsilon_{0} \varepsilon}{e d_{g}} \right), \nonumber
\end{align}
where $d$ and $d_{g}$ are the thickness of the tunnelling barrier and the distance between the drain and the back gate, respectively, $\tilde{\varepsilon}$ and $\varepsilon$ are the relative permittivities of the barrier and the substrate between the drain and gate and $n(\mu_{d})$ is the carrier density induced on the drain electrode. In sections III and IV, in order to demonstrate our general idea, we assume that the drain electrode consists of monolayer graphene, so that $n(\mu_{d}) = -\text{sgn}(\mu_d)\mu_{d}^{2}/\pi (v \hbar)^{2}$, with $v \approx 10^{6} \text{ms}^{-1}$, the Fermi velocity of the graphene electrons. We also assume that the tunnelling barrier is made of thin hexagonal boron nitride (hBN). As a result, the transmission coefficient $T$ in Eq. (\ref{CurrentEq}) only weakly depends on the energy of the initial state, $\epsilon$ \cite{britnell_sci_2012, georgiou_natnano_2013},  and so we set it to a constant (see appendix for a more detailed discussion). As a result, our current is not strictly provided in amperes but in arbitrary units - this, however, is enough to analyse NDR in the proposed devices.  Finally, the hBN located in the barrier is rotated by a large angle with respect to the graphene in the source/drain electrode. Because the impact of the hBN layer on graphene electrons decreases with increasing misalignment angle \cite{yankowitz_natphys_2012, jung_prb_2017}, this limits superlattice effects to those generated within the electrodes.

\section{Graphene on \lowercase{h}BN}

\begin{figure}
\subfloat{  \includegraphics[clip,width=1\columnwidth]{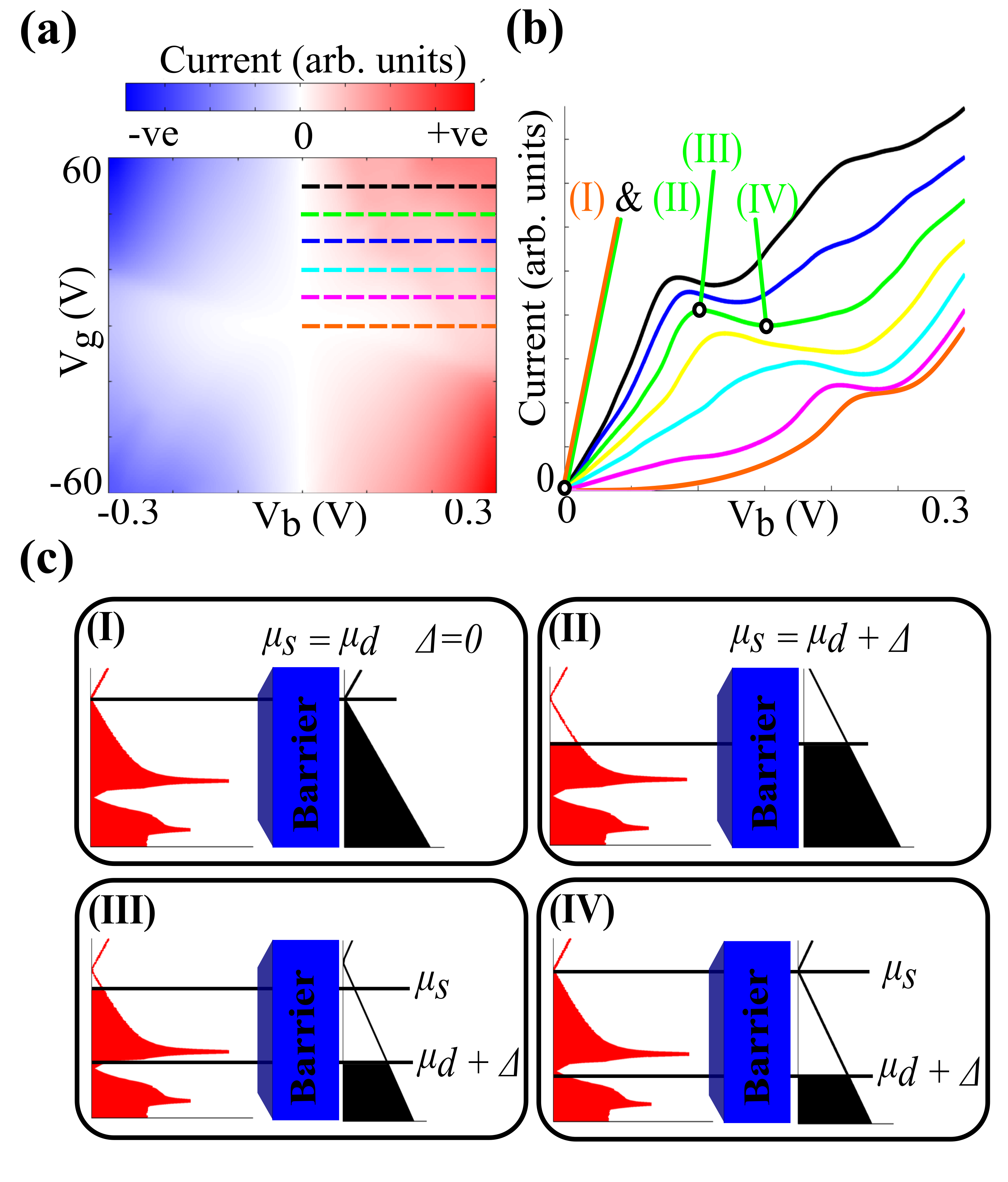} \label{fig:MLGhBNCurr}} 
\caption{(a) Calculated tunnelling current for a device from Fig. \ref{fig:device} as a function of voltages $V_{g}$ and $V_{b}$. (b) Tunnelling current as a function of $V_{b}$ for constant $V_{g}$ from 0 V (orange) to 50 V (black) in steps of 10 V. The cuts in ($V_{g}, V_{b}$) space corresponding to current curves in (b) are shown with dashed lines in (a). (c) Diagrams showing alignment of source and drain density of states as well as the positions of chemical potentials $\mu_{s}$ and $\mu_{d}+ \Delta$ for points in ($V_{g}, V_{b}$) space corresponding to tunnelling currents marked in (I), (II), (III) and (IV) in (b).}
\label{fig:MLGhBNCurrTotal}
\end{figure}

We first investigate the possibility of superlattice-induced NDR for a source electrode composed of hBN (layer 1) and monolayer graphene (layer 2). As the perturbing effect of hBN on graphene electrons decreases with increasing misalignment between the two crystals \cite{yankowitz_natphys_2012, jung_prb_2017}, we assume their crystalline axes are highly aligned. In such a case, the conical dispersion of graphene in the vicinity of the Brillouin zone (BZ) corner (valley) is folded into minibands, as indicated in Fig. 1b, with the valence band undergoing a more significant spectral reconstruction than the conduction band \cite{wallbank_physrevb_2013}, including the appearance of a Van Hove singularity (VHS) in the DOS \cite{wallbank_physrevb_2013, wang_natphys_2016, muchakruczynski_physrevb_2016}, shown in red in Fig. 1b.

In order to compute the density of states of the source electrode, we use an interlayer hopping model \cite{kindermann_prb_2012, wallbank_physrevb_2013, wallbank_physik_2015}, for the Hamiltonian of graphene electrons in the valley $\vect{K}_{\xi} = \xi \left(  4 \pi / 3a, 0 \right)$, $\xi = \pm 1$, perturbed by highly aligned hBN (assuming perfect alignment of the two crystals),

\begin{align}
    & \op{H}_\text{G-hBN} = \op{H}_{0} (\vect{p}, 0) + \boldsymbol{\delta} \op{H}, \label{MLGhBNEq}  \\
    & \boldsymbol{\delta} \op{H} = V_{0} \! \left( \frac{1}{2} f_{1}(\vect{r}) \! - \! \xi \frac{\sqrt{3}}{2}\sigma_{z} f_{2} (\vect{r}) \! - \! \dfrac{ \xi}{|\boldsymbol{b_0}|} [\op{l}_{z} \times \nabla f_{1}(\vect{r})] \! \cdot \! \vect{\sigma}\right),\nonumber  \\
    & f_1(\vect{r}) = \sum^5_{m=0} e^{i\vect{b_m}\cdot \vect{r}}, \quad f_2(\vect{r}) = i \sum^5_{m=0} (-1)^m e^{i\vect{b_m}\cdot \vect{r}} , \nonumber
\end{align}
written in the basis of $\{ \phi_{A, +}, \phi_{B, +} \}^{T}$ ( $\{ \phi_{B, -}, -\phi_{A, -} \}^{T}$) for $\xi= +1$ ($\xi=-1$) of Bloch states $\phi_{i, \xi}$ on one of the sublattices, $i = A, B$, that make up the graphene hexagons, calculated at the centre of the valley, $\vect{K}_\xi$. Also, we have introduced the Pauli matrices $\sigma_{x}$, $\sigma_{y}$ and $\sigma_{z}$, $\vect{\sigma} = (\sigma_{x}, \sigma_{y})$, acting in the sublattice space, $\vect{p} = (p_{x},p_{y})$ is the momentum of an electron as measured from the centre of the valley and $a$ is the lattice constant of graphene.

The first term in the Hamiltonian, $\op{H}_{0} (\vect{p}, \theta) = v e^{i \theta \sigma_z/2}\vect{\sigma} e^{-i \theta \sigma_z/2}\cdot \vect{p}$, describes the low energy, linear electronic dispersion of unperturbed graphene, while the term $\boldsymbol{\delta} \op{H}$ is due to the moir\'{e} potential with $k$-space periodicity given by a set of basic superlattice reciprocal vectors $\vect{b_m} = \op{R}_{\frac{2\pi m}{6}} \! \left[1- \op{R}_\theta /(1+\delta) \right] \! \left( 0, 4\pi /\sqrt{3}a \right)$, $m = 0, 1, ..., 5$, where $\op{R}_{\theta}$ is the anticlockwise rotation operator and $\delta = 1.8\%$ the lattice mismatch. The spatial variation of the superlattice potential is described by the two functions $f_{1}(\vect{r})$ and $f_{2}(\vect{r})$, linear combinations of the first harmonics of the moir\'{e}, and its strength is characterised by the parameter $V_{0}$, which we take equal to 17 meV \cite{lee_science_2016, plasmons_ni}.

In Fig. \ref{fig:MLGhBNCurr}, we present our simulation of the tunnelling current between the graphene/hBN source and graphene drain as a function of the voltages $V_b$ and $V_g$ in a device with hBN as the barrier ($d=1.3$ nm, $\tilde{\epsilon}=3$) and a Si gate separated by insulating layer ($d_g = 180$ nm, $\epsilon =3.9$), geometry similar to recent experiments \cite{ britnell_sci_2012, britnell_natcom_2013, lee_apl_2014, mishchenko_natnano_2014}. The appearance of NDR can be seen in the top right quarter of Fig. \ref{fig:MLGhBNCurr} where the tunnelling current decreases with increasing bias voltage. We show selected cuts through that region for various constant values of $V_{g}$ in Fig. 2b.

For $V_{b} = V_{g} = 0$ V, the chemical potentials in the source and drain are located at the respective neutrality points which are aligned with each other, as in diagram (I) in Fig. 2c, hence leading to an absence of tunnelling current. Applying the bias voltage introduces a relative shift between the source and drain chemical potentials $\mu_{s}$ and $\mu_{d}+\Delta$, respectively. As a result, an increase in $V_{b}$ for $V_{g} = 0$ V (corresponding to following the orange dashed line in Fig. \ref{fig:MLGhBNCurr} and 2b) leads to an increasing current as electrons from the valence band in the source can tunnel into the empty conduction band states of the drain. For $V_{b}$ slightly above 0.2 V, $\mu_{d}+\Delta$ moves past the moir\'{e}-induced VHS in the source valence band which leads to a shoulder-like feature in the orange curve in Fig. 2b.

In contrast, applying the gate voltage $V_{g}$ at constant $V_{b}$ dopes source and drain without affecting the energy difference between the chemical potentials, $\mu_{s}$ and $\mu_{d}+\Delta$. As shown in diagram (II) in Fig. 2c, for $V_{g} = 40$ V and $V_{b} = 0$ V no current flows through the structure because the chemical potentials are aligned, as in (I). Again, the current increases with increasing bias (as demonstrated by the green curve in Fig. 2b) until it reaches a peak when the occupied states in the moir\'{e}-induced VHS are aligned with empty states in the drain valence band, as in diagram (III). Because the VHS in the source DOS is followed by a dip, increasing $V_{b}$ further does not lead to more occupied electronic states contributing to the tunnelling. However, because changing $V_{b}$ affects the energy shift $\Delta$ between the Dirac points of the source and drain through Eq. (\ref{Electrostatics}), the number of empty states aligned with the VHS actually decreases with increasing $V_{b}$, as seen by comparing diagrams (III) and (IV). This results in a decrease of the current and NDR.

\section{Twisted Bilayer Graphene}

\begin{figure}
\subfloat{  \includegraphics[clip,width=\columnwidth,valign=t]{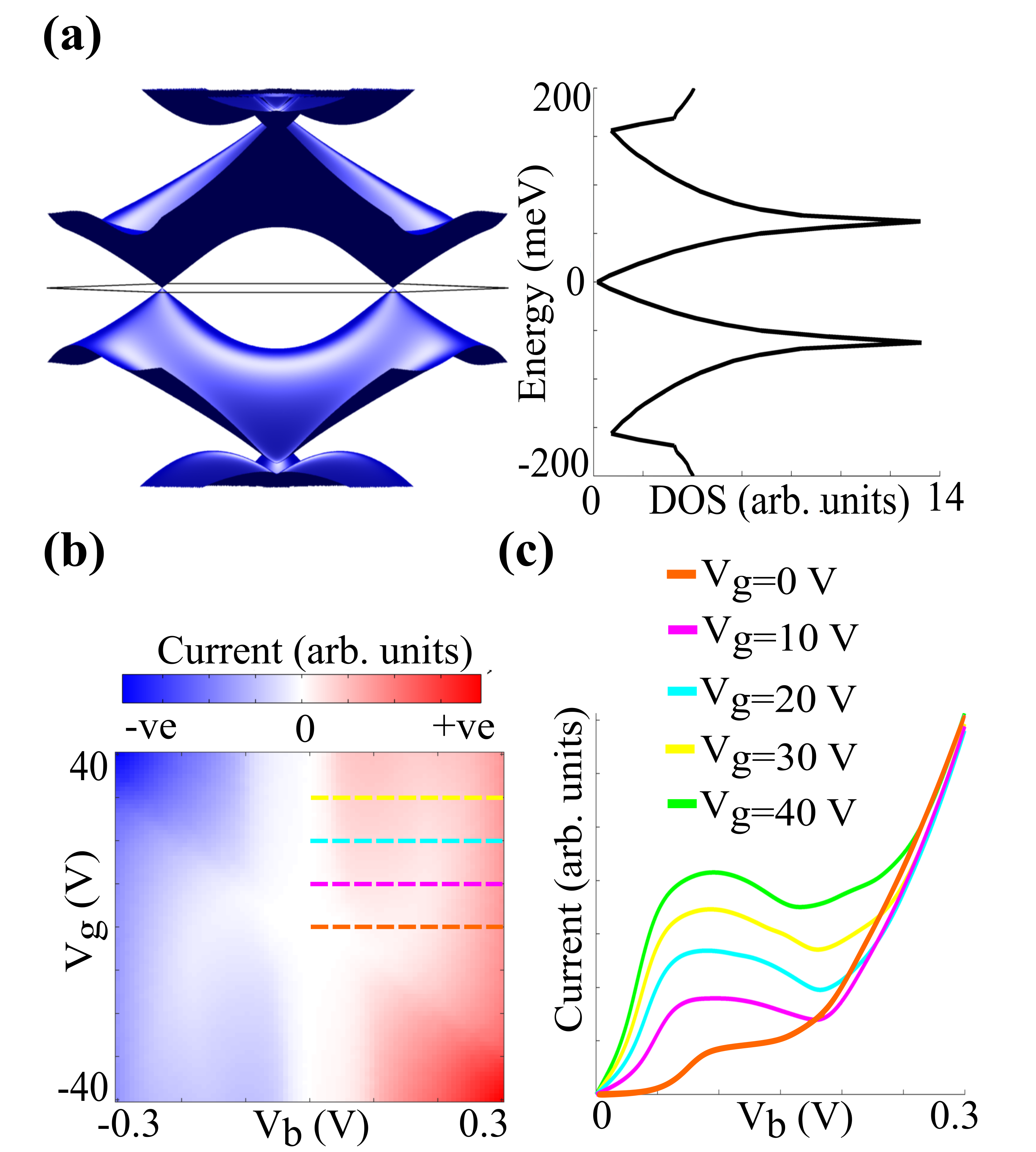} \label{2degBS}} 
\caption{(a) Low-energy band structure and density of states for twisted bilayer graphene  with a misalignment angle of $2^{\circ}$. Note that both the position and height of the VHS change with misalignment angle. The black line indicates the superlattice Brillouin zone edge. (b) Tunnelling current from tBLG to graphene across a hBN barrier as a function of gate and bias voltages $V_g$ and $V_b$. (c) Tunnelling current as a function of $V_b$ for constant $V_g$ from 0 V (orange) to 40 V (green) in steps of 10 V. The cuts in $(V_b, V_g)$ space corresponding to the current curves in (c) are shown with dashed lines in (b).}
\label{fig:2and3}
\end{figure}
To demonstrate the generality of our idea, we now discuss the existence of NDR in a twisted bilayer graphene/hBN/monolayer graphene Van der Waals tunnelling transistor. Twisted bilayer graphene \cite{li_natphys_2010, kkim_physrevlett_2012, robinson_acsnano_2013, campos_small_2013,cao_1, cao_2} (tBLG) comprises two stacked and rotationally misaligned graphene sheets. Because, in contrast to graphene on highly aligned hBN, in tBLG both layers have the same lattice constant, rotational misalignment is necessary to form a moir\'{e} superlattice. As a result of the twist, the interlayer coupling depends on the position $\vect{r}$ within the crystal, leading to an effective Hamiltonian \cite{santos_physrevlett_2007, bistritzer_pnas_2011, mele_physrevb_2011}

\begin{align}
    & \op{H}_\text{tBLG} = \begin{pmatrix}
 \op{H}_{0}(\vect{p},-\frac{\theta}{2}) + \dfrac{u}{2} & \op{T}^{\dagger} (\theta) \\
 \op{T} (\theta) & \op{H}_{0}(\vect{p},\frac{\theta}{2})- \dfrac{u}{2}
    \end{pmatrix}, \nonumber \\
 & \op{T} (\theta) = \frac{\gamma_{1}}{3} \sum_{j = 0}^{2} e^{i \hat{R}_{\frac{2 \pi j}{3}}\vect{\Delta K}\cdot \vect{r}} \begin{pmatrix} 
 1 & \xi e^{-i  j \frac{2 \pi}{3}} \\ \xi e^{i  j \frac{2 \pi}{3}} & 1
\end{pmatrix}, 
\label{tBLGHamiltonian}
\end{align}
where we included the effect of the electric field between the layers by introducing an on-site potential energy difference between the layers, $u$, related,  to the applied gate voltage
\begin{align}
    u  =  \dfrac{e^2 d_0}{\varepsilon_0} \left( n(\mu_d) + \frac{1}{2}n(\mu_s, u) + V_g \dfrac{\varepsilon_0 \varepsilon}{e d_g} \right) ,
\end{align}
where $d_0 = 0.33$nm is the bilayer graphene interlayer distance.
The Hamiltonian in Eq. (\ref{tBLGHamiltonian}), written in the basis of $\{ \phi_{A1,+}, \phi_{B1,+}, \phi_{A2,+}, \phi_{B2,+} \}^{T}$ ( $\{ \phi_{B1,-}, -\phi_{A1,-}, \phi_{B2,-}, -\phi_{A2,-} \}^{T}$) for $\xi= +1$ ($\xi=-1$), describes hybridisation of the two Dirac cones displaced by a vector $\Delta \vect{K} \approx 2 |\vect{K}_{\xi} |\sin \left( \theta/2 \right) (0,-1)$  from one another.  Midway between the cones, repulsion between two crossing linear dispersions leads to the opening of a local gap and appearance of a peak in the DOS, as shown in Fig. \ref{2degBS}. This peak has been observed by scanning tunnelling microscopy \cite{li_natphys_2010} and is known to modify the optical conductivity \cite{robinson_acsnano_2013, campos_small_2013} and Raman spectra \cite{kkim_physrevlett_2012} of tBLG. Unlike in Bernal-stacked bilayer graphene \cite{mccann}, $u$ does not open a band gap in tBLG but instead leads to small changes in the energy position of the  Dirac cones of each layer. We compute $u$ self-consistently assuming that the charge density is equally distributed between the two layers. The overall impact of $u$ on the DOS is small and all features in the I-V maps can be qualitatively explained with $u=0$.

For our modelling of the tunnelling between tBLG and graphene across a hBN multilayer, we choose the misalignment angle $2^\circ$, corresponding to the low-energy band structure in the vicinity of a single valley and density of states as shown in Fig. 3a. All the other geometrical parameters of the device are as used in the case of the graphene/hBN source electrode. The calculated current as a function of the bias and gate voltages $V_{b}$ and $V_{g}$ is shown in Fig. 3b and selected cuts for constant $V_{g}$ are presented in Fig. 3c. Similarly to the case of the graphene/hBN electrode, superlattice-induced spectral reconstruction, in particular the presence of sharp VHSs followed by a dip, leads to NDR for a range of gate voltages.

Because the VHS is a robust feature in the density of states of tBLG for a large range of misalignment angles \cite{brihuega_prl_2012}, the behaviour of the tunnelling current should also be similar for different $\theta$ (although note that greater misalignment angle requires higher $V_{g}$ to dope the source past the singularity). Also, because, in contrast to the aligned graphene/hBN heterostructure, density of states of tBLG is electron-hole symmetric, the graph in Fig. 3b is inversion-symmetric with respect to the point $V_{b} = V_{g} = 0$ V.

\section{Graphene on $\alpha - \text{In}_2 \text{Te}_2$}
While the architectures discussed in Sec. III and IV demonstrate the principle of moir\'{e}-induced NDR and are feasible experimentally, the calculated NDR peak-to-valley ratio is  only of order 1. It can be increased by choosing a different VdW heterostructure as an electrode, in particular, one with a moir\'{e} reconstructed density of states in which a bandgap between minigaps is close to a Van Hove singularity. Moreover, designing superlattices to modulate the densities of states of both the electrodes as opposed to using monolayer graphene with its linear DOS as a drain like in the two examples earlier, will also increase NDR.

Hence, in this section, we investigate current characteristics of a VdW tunneling transistor with both electrodes made of graphene on $\alpha - \text{In}_2 \text{Te}_2$. It was proposed that such a heterostructure would belong to a group for which the moir\'{e} pattern results from the beating between the lattice constants of $\alpha - \text{In}_2 \text{Te}_2$ and that of a tripled graphene unit cell, $\sqrt{3}a$ \cite{wallbank_physrevb_2013_root3, leech_physrevb_2016}, with the mismatch between the two approximately $\delta' \approx -0.7\%$ \cite{zolyomi_2014, ortix}. It was predicted \cite{wallbank_physrevb_2013_root3}, in this case, that the arising moir\'{e} potential would lead to a periodically oscillating in space intervalley coupling for graphene electrons, captured by the Hamiltonian,
\begin{align}
 & \op{H}_\text{G-$\sqrt{3}$} = \op{H}_{0} (\vect{p}, 0) + \boldsymbol{\delta} \op{H}', \label{MLGhBNEq}  \\
    & \boldsymbol{\delta} \op{H}' = V'_{0} \! \left( \frac{1}{2} F(\boldsymbol{r}) -\dfrac{1}{|\boldsymbol{\beta}_0|}[\boldsymbol{\sigma} \times \op{l}_z]\cdot \nabla F(\boldsymbol{r})\right),\nonumber  \\
    & f_1(\vect{r}) = \sum^5_{m=0} e^{i\vect{\beta_m}\cdot \vect{r}}, \quad f_2(\vect{r}) = i \sum^5_{m=0} (-1)^m e^{i\vect{\beta_m}\cdot \vect{r}} ,\nonumber 
    \\ &F(\boldsymbol{r})  = \tau_x f_1(\vect{r}) + \tau_y f_2(\vect{r}), \nonumber
\label{root3Hamiltonian}
\end{align}
written in the basis $\{ \phi_{A, +}, \phi_{B, +}, \phi_{B, -}, \phi_{-A, -} \}^{T}$ using Pauli matrices $\tau_x$, $\tau_y$ acting in the valley space and a set of reciprocal space vectors $\vect{\beta_m}  =\frac{1}{\sqrt{3}} \op{R}_{\frac{-\pi}{2}} \vect{b_m} $, $m=0,1,...,5$. Just like for graphene on hBN, the term $\boldsymbol{\delta} \op{H}'$ describes the contribution due to the moir\'{e} potential with $V'_0$ setting its strength. We assume that the interlayer interactions in graphene/$\alpha - \text{In}_2 \text{Te}_2$ are comparable to graphene/hBN and so use $V'_0 = V_0 =17$ meV for our calculations.   The calculated miniband spectrum is shown in Fig. 4a. Notice that, in contrast to the miniband spectra discussed for graphene on hBN in Sec. II, this one displays flat bands around zero energy, similar to magic-angle twisted bilayer graphene \cite{cao_1, cao_2}, with small band gaps separating them from the rest of the spectrum. This leads to a DOS as shown in Fig. 4b, containing Van Hove singularities next to a window of zero density of states, features attractive for an increased NDR. In Fig. 4c, we present the current calculated as a function of the gate and bias voltages for a VdW tunneling transistor incorporating the graphene/$\alpha - \text{In}_2 \text{Te}_2$ heterostructures as both source and drain (all other parameters of the device are kept the same as in Sec. II and III). Current curves for selected constant gate voltages and changing bias are shown in Fig. 4d (region of positive $V_g$ and $V_b$) and Fig. 4e (region of negative $V_g$ and $V_b$). For easier comparison with other figures in this paper, we have reversed the current and bias voltage axes in Fig. 4e. Moreover the current scale in Fig. 4d has been scaled by a factor of 3 as compared to Fig. 4e.


All of the I-V characteristics in Fig. 4d and 4e show NDR peak-to-valley ratios ranging from 2 to 10, depending on the choice of gate voltage. The largest NDR of around 10 is that for $V_g = 0$ V (orange curve) in Fig. 4d. It results from the presence of two Van Hove singularities in the DOS around zero energy (see Fig. 4b) and their movement on the energy scale in the source/drain electrodes as a function of $V_b$.
\begin{figure}
\subfloat{  \includegraphics[clip,width=\columnwidth]{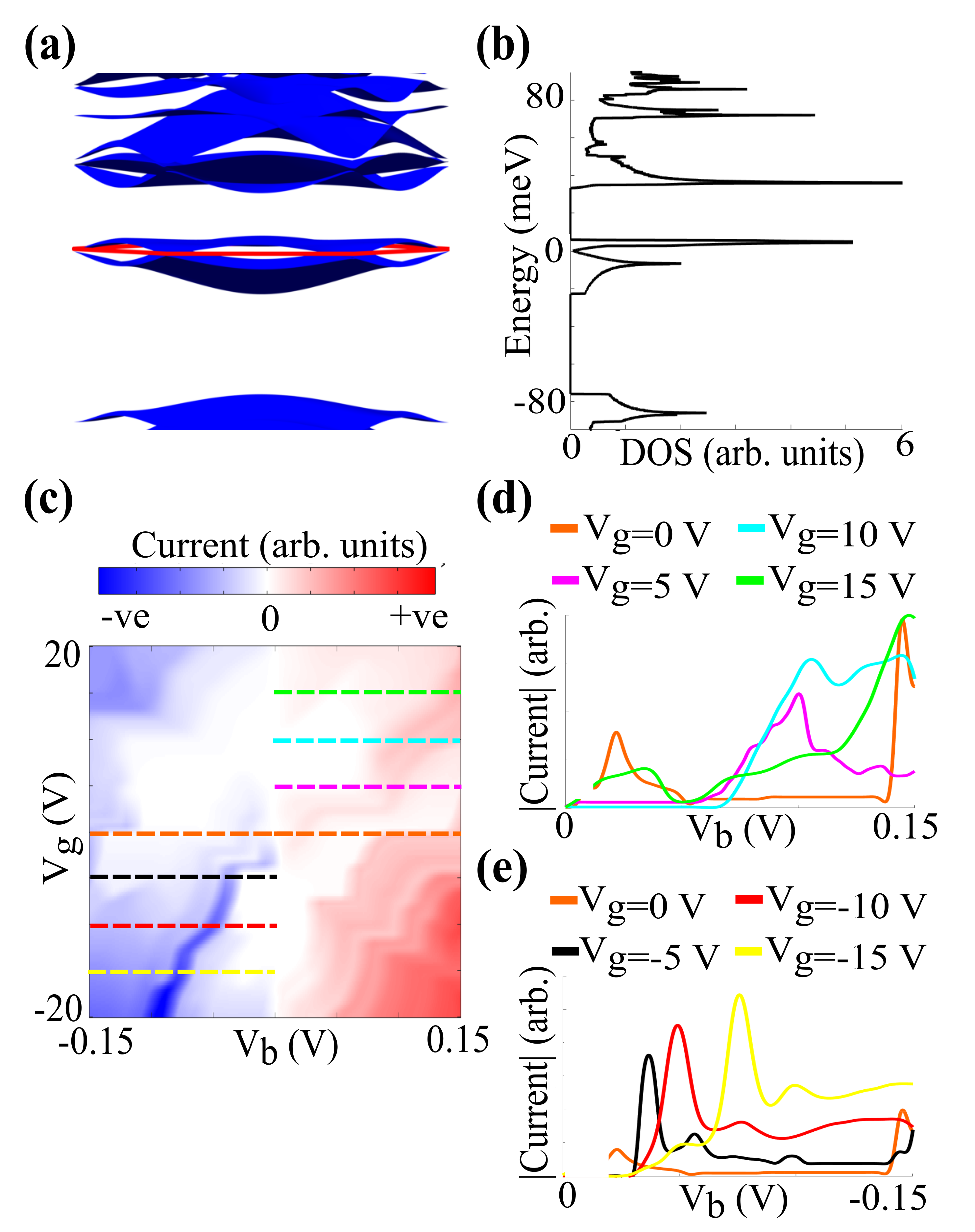} \label{root31}}
\caption{(a) Low-energy band structure of graphene on aligned $\alpha - \text{In}_2\text{Te}_2$ . The outline of the superlattice Brillouin zone is shown in red for clarity. (b) Corresponding density of states. (c) Tunnelling current between two graphene on aligned $\alpha - \text{In}_2\text{Te}_2$ electrodes across a hBN barrier as a function of gate and bias voltages $V_g$ and $V_b$. (d) Absolute  tunnelling current as a function of $V_b$ for constant $V_g$ from 0 V (orange) to 15 V (green) in steps of 5 V. The cuts in $(V_b, V_g)$ space corresponding to the current curves in (d) are shown with dashed lines in (c).  The current lines in (d) are scaled to highlight NDR features. (e) Same as (d) except for negative gate and bias voltages with steps of -5 V from 0 V (orange) to -15 V (yellow). }
\label{fig:4}
\end{figure}

\section{Summary}

In summary, we demonstrated theoretically for three different architectures that the modifications of the electronic density of states due to the formation of moir\'{e} superlattices of Van der Waals crystals can lead to negative differential resistance when the moir\'{e} heterostructure is employed as an electrode in a vertical tunnelling transistor. This is achieved without the requirement of momentum-conserving tunnelling, which has only been observed in the highest quality, closely-aligned, devices, made of mechanically exfoliated crystals. For this reason, our idea might be useful for materials produced by other methods like chemical vapour deposition, where clear moir\'{e} reconstruction has been observed \cite{li_natphys_2010, PhysRevB.92.155409} but no momentum-conserving tunnelling has been reported. While the moir\'{e} superlattices used in the first two examples, graphene on hBN and twisted bilayer graphene, have been realised experimentally, the last one, graphene on $\alpha - \text{In}_2\text{Te}_2$, has not. However, superlattice effects have been observed or predicted for a variety of different heterostructures and interfaces \cite{ leech_physrevb_2016, tong_natphys_2017, zhang_sciadv_2017, pan_nanolett_2018}, so that significant NDR peak-to-valley ratio might indeed be possible for certain architectures, as suggested in Sec. V. Importantly, in contrast to artificial superlattices, our idea avoids the need to process any of the two-dimensional crystals after they are grown, as the superlattice is provided by an interface between neighbouring layers. This, in turn, limits disorder and degradation of the components, especially important if the starting materials were not obtained by mechanical exfoliation. 
In contrast to many other NDR setups, our idea is, by design, easy to integrate in more complicated devices based on two-dimensional crystals and VdW interfaces. It can also be coupled with artificial patterning of dielectric substrates underneath 2D materials on length scales comparable to moir\'{e} wavelengths \cite{forsythe_arxiv_2018}.

\begin{acknowledgements}

This work has been supported by the EPSRC through the University of Bath Doctoral Training Partnership, grant numbers EP/M50645X/1 and EP/M507982/1.

Authors D.J.L. and J.J.P.T. contributed equally to this work.
\end{acknowledgements}

\appendix

\section{Energy Dependence of Transmission Coefficient}

For thin, strongly insulating tunnelling barriers (such as hBN), the transmission coefficient $T$  only weakly depends on the energy of the initial state, $\epsilon$ \cite{britnell_sci_2012, georgiou_natnano_2013},  and so we set it to a constant in the main text for ease of analysis.
In general, the variation in the barrier height $\Phi(\epsilon)$ due to  changing electron kinetic energy leads to a modification of the  decay constant $k$ which becomes a function of $\epsilon$ and so does the transmission coefficient, which decreases exponentially with barrier thickness, $d$,

\begin{align}
    T(\epsilon) = \exp[- k(\epsilon) d].
\end{align}
 Assuming graphene electrodes, the multilayer hBN barrier can be treated as an isotropic potential step  \cite{britnell_sci_2012} with barrier height $\Phi_0 = -1.5$ eV, corresponding to previous measurements of the valence band maximum (VBM) of hBN  \cite{britnell_sci_2012, georgiou_natnano_2013}. Approximating hBN energy bands as roughly parabolic around the VBM allows us to write \cite{simmons_1963}

\begin{align}
    k(\epsilon) = \text{Im} \dfrac{\sqrt{2 m^* \Phi(\epsilon)}}{\hbar} =  \text{Im} \dfrac{\sqrt{2 m^* (\Phi_0-\epsilon)}}{\hbar} , 
\end{align}
where $m^*\approx 0.5m_0$ is the effective mass \cite{britnell_sci_2012, georgiou_natnano_2013}. Notably, this predicts electron-hole asymmetry in tunnelling current as observed in experiment \cite{britnell_sci_2012, georgiou_natnano_2013}. 

Applying this varying tunnelling coefficient to our model increases the tunnelling current and conductance in the conduction band-conduction band voltage regions and reduces the current and conductance in the valence band-valence band regions. Crucially, all NDR features shown in our calculations still persist, with conduction band peak-to-valley ratio slightly increased and valence band peak-to-valley slightly decreased in twisted bilayer, graphene on hBN and double graphene/$\alpha - \text{In}_2 \text{Te}_2$. Increasing the tunnelling barrier thickness or using a less insulating material such as W$\text{S}_2$ \cite{georgiou_natnano_2013} would require the potential modulation of the transmission coefficient to be included in order to give accurate results. However, despite the observation that across the investigated voltage range the transmission coefficient varies noticeably, locally, around the current peak and valley voltages, the transmission coefficient is roughly constant and so all NDR features that we predict will persist. 

In general formalism, determining the tunnelling current requires finding the overlap between the relevant electron wave functions on the electrodes. This can be decomposed into a term describing the transverse component of the overlap,  $T(\epsilon)$, and an in-plane momentum-conserving component, which is constant for momentum non-conserving tunnelling. In real devices, the total current is a sum of all tunnelling processes, both conserving and not conserving momentum. However as mentioned in the main text, the upper bounds on the current are set by the available initial/final DOS so that any contributions not considered here cannot be larger than those we discuss in the text. The arbitrariness of our units of current in Fig. 2-4 originates in setting $T(\epsilon)$ to an unspecified constant which affects any other tunnelling process in the same way. 

\bibliographystyle{unsrt}

\begin{thebibliography}{99}

\bibitem{esaki_physrev_1958} L. Esaki, New phenomenon in narrow germanium para-normal-junctions, Phys. Rev. \textbf{109}, 603 (1958).

\bibitem{britnell_natcom_2013} L. Britnell, R. V. Gorbachev, A. K. Geim, L. A. Ponomarenko, A. Mishchenko, M. T. Greenaway, T. M. Fromhold, K. S. Novoselov, and L. Eaves, Resonant tunnelling and negative differential conductance in graphene transistors, Nat. Comms. \textbf{4}, 1794 (2013).

\bibitem{mishchenko_natnano_2014} A. Mishchenko, J. S. Tu, Y. Cao, R. V. Gorbachev, J. R. Wallbank, M. T. Greenaway, V. E. Morozov, S. V. Morozov, M. J. Zhu, S. L. Wong, F. Withers, C. R. Woods, Y. J. Kim, K. Watanabe, T. Taniguchi, E. E. Vdovin, O. Makarovsky, T. M. Fromhold, V. I. Fal'ko, A. K. Geim, L. Eaves, and K. S. Novoselov, Twist-controlled resonant tunnelling in graphene/boron nitride/graphene heterostructures, Nat. Nano. \textbf{9}, 808 (2014).
 

\bibitem{fallahazad_nanolett_2015}B. Fallahazad, K. Lee, S. Kang, J. M. Xue, S. Larentis, C. Corbet, K. Kim, H. C. P. Movva, T. Taniguchi, K. Watanabe, L. F. Register, S. K. Banarjee, and E. Tutuc, Gate-Tunable Resonant Tunneling in Double Bilayer Graphene Heterostructures, Nano Lett. \textbf{15}, 428 (2015).


\bibitem{kang_ieee_2015} S. Kang, B. Fallahazad, K. Lee, H. Movva, K. Kim, C. M. Corbet, T. Taniguchi, K. Watanabe, L. Colombo, L. F. Register, E. Tutuc, and S. K. Banerjee, Bilayer graphene–hexagonal boron nitride heterostructure negative differential resistance interlayer tunnel FET, IEEE Electron Device Lett. \textbf{36}, 405 (2015).


\bibitem{kkim_nanolett_2016}  K. Kim, M. Yankowitz, B. Fallahazad, S. Kang, H. C. P. Movva, S. Q. Huang, S. Larentis, C. M. Corbet, T. Taniguchi, K. Watanabe, S. K. Banerjee, B. J. LeRoy, and E. Tutuc, Van der Waals Heterostructures with high accuracy rotational alignment, Nano Lett. \textbf{16}, 1989 (2016).


\bibitem{wallbank_sci_2016}      J. R. Wallbank, D. Ghazaryan, A. Misra, Y. Cao, J. S. Tu, B. A. Piot, M. Potemski, S. Pezzini, S. Wiedmann, U. Zietler, T. L. M. Lane, S. V. Morozov, M. T. Greenaway, L. Eaves, A. K. Geim, V. I. Fal'ko, K. S. Novoselov, and A. Mishchenko, Tuning the valley and chiral quantum state of Dirac electrons in Van der Waals heterostructures, Science \textbf{353}, 575 (2016).


\bibitem{haigh_natmat_2012} S. J. Haigh, A. Gholinia, R. Jalil, S. Romani, L. Britnell, D. C. Elias, K. S. Novoselov, L. A. Ponomarenko, A. K. Geim, and R. Gorbachev,Cross-sectional imaging of individual layers, buried interfaces of graphene-based heterostructures, superlattices, Nat. Mat. \textbf{11}, 764 (2012).


\bibitem{lee_apl_2014}     S. H. Lee, M. S. Choi, J. Lee, C. H. Ra, X. Liu, E. Hwang, J. H. Choi, J. Q. Zhong, W. Chen, and W. J. Yoo, High performance vertical tunneling diodes using graphene/hexagonal boron nitride/graphene hetero-structure, Appl. Phys. Lett. \textbf{104}, 053103 (2014).


\bibitem{roy_applphyslett_2014} T. Roy, L. Liu, S. de la Barrera, B. Chakrabarti, Z. R. Hesabi, C. A. Joiner, R. M. Feenstra, G. Gu, and E. M. Vogel, Tunneling characteristics in chemical vapor deposited graphene-hexagonal boron nitride-graphene junctions, Appl. Phys. Lett. \textbf{104}, 123506 (2014).

\bibitem{yankowitz_natphys_2012}  M. Yankowitz, J. M. Xue, D. Cormode, J. D. Sanchaz-Yamagishi, K. Watanabe, T. Taniguchi, P. Jarillo-Herrero, P. Jacquod, and B. J. LeRoy, Emergence of superlattice Dirac points in graphene on hexagonal boron nitride, Nat. Phys. \textbf{8}, 382 (2012).


\bibitem{wang_natphys_2016} E. Y. Wang, X. B. Lu, S. J. Ding, W. Yao, M. Z. Yan, G. L. Wan, K. Deng, S. P. Wang, G. R. Chen, L. G. Ma, J. Jung, A. V. Fedorov, Y. B. Zhang, G. Y. Zhang, S. Y. Zhou, Gaps induced by inversion symmetry breaking and second-generation Dirac cones in graphene/hexagonal boron nitride, Nat. Phys. \textbf{12}, 1111 (2016).

\bibitem{shi_natphys_2014} Z. W. Shi, C. H. Jin, W. Yang, L. Ju, J. Horng, X. B. Lu, H. A. Bechtel, M. C. Martin, D. Y. Fu, J. Q. Wu, K. Watanabe, T. Taniguchi, Y. B. Zhang, X. D. Bai, E. G. Wang, G. Y. Zhang, and F. Wang, Gate-dependent pseudospin mixing in graphene/boron nitride moire superlattices, Nat. Phys. \textbf{10}, 743 (2014). 

\bibitem{yang_nanolett_2016} W. Yang, X. Lu, G. Chen, S. Wu, G. B. Xie, M. Cheng, D. M. Wang, R. Yang, D. X. Shi, K. Watanabe, T. Taniguchi, C. Voisin, B. Placais, Y. B. Zhang, and G. Y. Zhang, Hofstadter Butterfly and Many-Body Effects in Epitaxial Graphene Superlattice, Nano Lett. \textbf{16}, 2387-2392 (2016).

\bibitem{li_natphys_2010} G. H. Li, A. Luican, J. M. B. L. dos Santos, A. H. Castro Neto, A. Reina, J. Kong, and E. Y. Andrei, Observation of Van Hove singularities in twisted graphene layers, Nat. Phys. \textbf{6}, 109-113 (2010).

\bibitem{yan_physrevlett_2012} W. Yan, M. X. Liu, R. F. Dou, L. Meng, L. Feng, Z. D. Chu, Y. F. Zhang, Z. F. Liu, J. C. Nie, and L. He, Angle-Dependent Van Hove Singularities in a Slightly Twisted Graphene Bilayer, Phys. Rev. Lett. \textbf{109}, 126801 (2012).

\bibitem{tan_acsnano_2016} Z. J. Tan, J. Yin, C. Chen, H. Wang, L. Lin, L. Z. Sun, J. X. Wu, X. Sun, H. F. Yang, Y. L. Chen, H. L. Peng, and Z. F. Liu, Building Large-Domain Twisted Bilayer Graphene with Van Hove Singularity, ACS Nano \textbf{10}, 6725-6730 (2016).

\bibitem{ykim_nanolett_2016} Y. Kim, P. Herlinger, P. Moon, M. Koshino, T. Taniguchi, K. Watanabe, and J. H. Smet, Charge Inversion and Topological Phase Transition at a Twist Angle Induced Van Hove Singularity of Bilayer Graphene, ACS Nano \textbf{16}, 5053-5059 (2016).

\bibitem{pierucci_nanolett_2016} D. Pierucci, H. Henck, J. Avila, A. Balan, C. H. Naylor, G. Patriarche, Y. J. Dappe, M. G. Silly, F. Sirotti, A. T. C. Johnson, M. C. Asensio, and A. Ouerghi, Band Alignment and Minigaps in Monolayer MoS2-Graphene van der Waals Heterostructures, Nano Lett. \textbf{16}, 4054-4061 (2016).

\bibitem{koren_natnanotech_2016} E. Koren, I. Leven, E. Lortscher, A. Knoll, O. Hod, and U. Duerig, Coherent commensurate electronic states at the interface between misoriented graphene layers, Nat. Nano. \textbf{11}, 752-757 (2016).

\bibitem{brihuega_prl_2012} I. Brihuega, P. Mallet, H. Gonzalez-Herrero, G. T. de Laissardiere, M. M. Ugeda, L. Magaud, J. M. Gomez-Rodriguez, F. Yndurain, and J. Y. Veuillen, Unraveling the Intrinsic and Robust Nature of Van Hove Singularities in Twisted Bilayer Graphene by Scanning Tunneling Microscopy and Theoretical Analysis, Phys. Rev. Lett. \textbf{109}, 196802 (2012).

\bibitem{lee_science_2016} M. Lee, J. R. Wallbank, P. Gallagher, K. Watanabe, T. Taniguchi, V. I. Fal'ko, and D. Goldhabor-Gordon, Ballistic miniband conduction in a graphene superlattice, Science \textbf{353}, 1526-1529 (2016).

\bibitem{moffat} P. Moffatt, and E. H. Kim, Negative differential resistance from a Van Hove singularity in tunnel diodes, Appl. Phys. Lett. \textbf{89} , 192117 (2006).


\bibitem{britnell_sci_2012} L. Britnell, R. V. Gorbachev, R. Jalil, B. D. Belle, F. Schedin, A. Mishchenko, T. Georgiou, M. I. Katnelson, L. Eaves, S. V. Morozov, N. M. R. Peres, J. Leist, A. K. Geim, K. S. Novoselov, and L. A. Ponomarenko, Field-effect tunneling transistor based on vertical graphene heterostructures, Science \textbf{335}, 947-950 (2012).

\bibitem{georgiou_natnano_2013} T. Georgiou, R. Jalil, B. D. Belle, L. Britnell, R. V. Gorbachev, S. V. Morozov, Y. J. Kim, A. Gholinia, S. J. Haigh, O. Makarovsky, L. Eaves, L. A. Ponomarenko, A. K. Geim, K. S. Novoselov, and A. Mishchenko, Vertical field-effect transistor based on graphene-WS$_{2}$ heterostructures for flexible and transparent electronics, Nat. Nano. \textbf{8}, 100-103 (2013).

\bibitem{jung_prb_2017} J. Jung, E. Laksono, A. M. DaSilva, A. H. MacDonald, M. Mucha-Kruczy\'{n}ski, and S. Adam, Moir\'{e} band model and band gaps of graphene on hexagonal boron nitride, Phys. Rev. B \textbf{96}, 085442 (2017).

\bibitem{wallbank_physrevb_2013} J. R. Wallbank, A. A. Patel, M. Mucha-Kruczy\'{n}ski, A. K. Geim, and V. I. Fal'ko, Generic Miniband Structure of Graphene on a Hexagonal Substrate, Phys. Rev. B \textbf{87}, 245408 (2013).



\bibitem{muchakruczynski_physrevb_2016} M. Mucha-Kruczy\'{n}ski, J. R. Wallbank and V. I. Fal'ko, Moir\'{e} miniband features in the angle-resolved photoemission spectra of graphene/hBN heterostructures, Phys. Rev. B \textbf{93}, 085409 (2016).

\bibitem{wallbank_physik_2015} J. R. Wallbank, M. Mucha-Kruczy\'{n}ski, X. Chen, and V. I. Fal'ko, Moir\'{e} miniband features in the angle-resolved photoemission spectra of graphene/hBN heterostructures, Annalen Der Physik \textbf{527}, 359-376 (2015).








\bibitem{kindermann_prb_2012} M. Kindermann, B. Uchoa, and D. L. Miller, Zero-energy modes and gate-tunable gap in graphene on hexagonal boron nitride, Phys. Rev. B \textbf{86}, 115415 (2012).

\bibitem{plasmons_ni} G. X. Ni, H. Wang, J. S. Wu, Z. Fei, M. D. Goldflam, F. Keilmann, B. Özyilmaz, A. H. Castro Neto, X. M. Xie, M. M. Fogler and D. N. Basov, Plasmons in graphene moiré superlattices, Nature Materials \textbf{14},1217-–1222 (2015).

\bibitem{santos_physrevlett_2007} J. M. B. Lopes dos Santos, N. M. R. Peres, and A. H. Castro, Graphene bilayer with a twist: Electronic structure, Phys. Rev. Lett. \textbf{99}, 256802 (2007).

\bibitem{bistritzer_pnas_2011} R. Bistritzer, and A. H. MacDonald, Moir\'{e} bands in twisted double-layer graphene, Proc. Nat. Acad. Sci. USA \textbf{108}, 12233--12237 (2011).

\bibitem{mele_physrevb_2011} E. J. Mele, Band symmetries and singularities in twisted multilayer graphene, Phys. Rev. B \textbf{84}, 235439, (2011).

\bibitem{kkim_physrevlett_2012} K. Kim, S. Coh, L. Z. Tan, W. Regen, J. M. Yuk, E. Chatterje, M. F. Crommie, M. L. Cohen, S. G. Louie, and A. Zettl, Raman Spectroscopy Study of Rotated Double-Layer Graphene: Misorientation-Angle Dependence of Electronic Structure, Phys. Rev. Lett. \textbf{108}, 246103 (2012).

\bibitem{robinson_acsnano_2013} J. T. Robinso, S. W. Schmucker, C. B. Diaconescu, J. P. Long, J. C. Culbertson, T. Ohta, A. L. Friedman, and T. E. Beechem, Electronic Hybridization of Large-Area Stacked Graphene Films, ACS Nano \textbf{7}, 637-644 (2013).

\bibitem{campos_small_2013} J. Campos-Delgado, G. Algara-Siller, C. N. Santos, U. Kaiser, and J. P. Raskin, Twisted Bi-Layer Graphene: Microscopic Rainbows, Small \textbf{9}, 3247-3251 (2013).



\bibitem{cao_1} Y. Cao, V. Fatemi, A. Demir, S. Fang, S. L. Tomarken, J. Y. Luo, J. D. Sanchez-Yamagishi, K. Watanabe, T. Taniguchi, E. Kaxiras, R. C. Ashoori and P. Jarillo-Herrero, Correlated insulator behaviour at half-filling in magic-angle graphene superlattices, Nature \textbf{556}, 80–-84 (2018).

\bibitem{cao_2} Y. Cao, V. Fatemi, S. Fang, K. Watanabe, T. Taniguchi, E. Kaxiras and P. Jarillo-Herrero, Unconventional superconductivity in magic-angle graphene superlattices, Nature \textbf{556}, 43–-50 (2018).

\bibitem{mccann} E. McCann, V. I. Fal'ko,  Landau-Level Degeneracy and Quantum Hall Effect in a Graphite Bilayer, Phys. Rev. Lett. \textbf{96}, 086805 (2006).

\bibitem{wallbank_physrevb_2013_root3} J. R. Wallbank, M. Mucha-Kruczy\'{n}ski and V. I. Fal'ko, Moir\'{e} minibands in graphene heterostructures with almost commensurate $\sqrt{3} \times \sqrt{3}$ hexagonal crystals, Phys. Rev. B \textbf{88}, 155415 (2013).

\bibitem{leech_physrevb_2016} D. J. Leech, and M. Mucha-Kruczy\'{n}ski, Controlled formation of an isolated miniband in bilayer graphene on an almost commensurate $\sqrt{3} \times \sqrt{3}$ substrate, Phys. Rev. B \textbf{94}, 165437 (2016).

\bibitem{zolyomi_2014} V. Zolyomi, N. D. Drummond and V. I. Fal'ko, Electrons and Phonons in Single Layers of Hexagonal Indium Chalcogenides from Ab Initio Calculations, Phys. Rev. B \textbf{89},  205416 (2014).

 \bibitem{ortix} G. Giovannetti, M. Capone, J. van den Brink and C.Ortix, Kekul\'e textures, pseudospin-one Dirac cones, and quadratic band crossings in a graphene-hexagonal indium chalcogenide bilayer,  Phys. Rev. B \textbf{91}, 121417 (2015).
 
\bibitem{PhysRevB.92.155409} D. Wong, Y. Wang, J. Jung, S. Pezzini, A. M. DaSilva, H. Tsai, H. S. Jung, R. Khajeh, Y. Kim, J. Lee, S. Kahn, S. Tollabimazraehno, H. Rasool, K. Watanabe, T. Taniguchi, A. Zettl, S. Adam, A. H. MacDonald and M. F. Crommie, Local spectroscopy of moir\'e-induced electronic structure in gate-tunable twisted bilayer graphene, Phys. Rev. B. \textbf{92}, 155409 (2015).

\bibitem{tong_natphys_2017} Q. Tong, H. Y. Yu, Q. Z. Zhu, Y. Wang, X. D. Xu, and A. Yao, Topological mosaics in moir\'{e} superlattices of van der Waals heterobilayers, Phys. Rev. Lett. \textbf{13}, 356-362 (2017).


\bibitem{zhang_sciadv_2017} C. D. Zhang, C. P. Chuu, X. B. Ren, L. J. Li, C. H. Jin, M. Y. Chou, and C. K. Shih, Interlayer couplings, Moir\'{e} patterns, and 2D electronic superlattices in MoS$_{2}$/WSe$_{2}$ hetero-bilayers, Sci. Adv. \textbf{3}, 1601459 (2017).

\bibitem{pan_nanolett_2018} Y. Pan, S. F\"{o}lsch, Y. Nie, D. Waters, Y. C. Li, B. Jariwala, K. Zhang, K. Cho, J. A. Robinson and R. M. Feenstra, Quantum-Confined Electronic States Arising from the Moir\'{e} Pattern of MoS$_{2}$-–WSe$_{2}$ Heterobilayers, Nano Lett. \textbf{18}, 1849-1855 (2018).

\bibitem{forsythe_arxiv_2018} C. Forsythe, X. Zhou, T. Taniguchi, K. Watanabe, A. Pasupathy, P. Moon, M. Koshino, P. Kim, and C. R. Dean, Band structure engineering of 2D materials using patterned dielectric superlattices, Nat. Nano. (2018) DOI:10.1038/s41565-018-0138-7. 

\bibitem{simmons_1963} J. G. Simmons, Formula for the electric tunnel effect between similar electrodes separated by a thin insulating film, Jour. App. Phys. \textbf{34}, 6 (1963).

\end{thebibliography}

\end{document}